\documentclass[12pt,preprint]{aastex}
\usepackage{amsmath, amsthm, amssymb}

  \def\ltsima{$\; \buildrel  <  \over \sim
  \;$}    \def\simlt{\lower.5ex\hbox{\ltsima}}    
\def\gtsima{$\;      \buildrel      >      \over      \sim      \;$}
\def\simgt{\lower.5ex\hbox{\gtsima}}      

\begin{document} 
   
\title{The Transition from ``Normal'' to ``Broad Absorption Line
  Quasar'' of Ton 34}
\author{Krongold, Y.\altaffilmark{1}; Binette, L.\altaffilmark{1};
  Hern\'andez-Ibarra, F.\altaffilmark{1}}
\altaffiltext{1}{Instituto de Astronomia, Universidad Nacional
  Autonoma de Mexico, Apartado Postal 70-264, 04510 Mexico DF,
  Mexico.}


\begin{abstract} 
We report the emergence of a high velocity,
broad absorption line
outflow in the luminous quasar Ton 34, at z$_q$=1.928. The outflow is
detected through an ultraviolet CIV broad absorption line, in a
spectrum  obtained in January 2006 by the Sloan Digital Sky
Survey. No absorption trough was present in two different
spectra acquired in 1981 at Las Campanas and Palomar observatories,
indicating the emergence of the outflow in less
than $\sim$8 yr (rest-frame). The absorption line spans a velocity range
from $\sim$5,000-25,000
km s$^{-1}$, and resembles typical troughs found in Broad Absorption
Line quasars (BALQSOs). We measure a balnicity index $\simgt$600
(though this value might be an underestimation due to a conservative
placing of the continuum). The absorption trough is likely
saturated, with the absorbing gas covering $\sim$25\% of the emitting
region. We explore different scenarios for the
emergence of this outflow, and find an existing wind moving across our
line of sight to the source as the most likely explanation. This
indicates that high velocity outflows (producing broad absorption
troughs in BALQSOs) might be ubiquitous in quasars, yet only become
observable when the wind accidentally 
crosses our line vision to the central source.       

\end{abstract}


\section{Introduction \label{par:intro}}
More than 50\% of active galaxies and quasars display clear 
signatures of outflows, easily identified in the form of
absorption lines (e.g. Crenshaw et al. 2003; Ganguly \& Brotherton
2008). These lines appear in active galactic nuclei (AGN)
spectra in different forms, either as broad absorption lines (BALs;
FWHM$\sim$10,000 km s$^{-1}$), narrow absorption lines (NALs;
FWHM $\sim$100 km s$^{-1}$) and intermediate mini-BALs. 

Understanding these winds is essential to understand the
structure and dynamics of quasars (Elvis 2000). Since they carry mass
and kinetic energy outside
the central region of the host galaxy, it has been suggested
that they may also be essential for galaxy evolution, being responsible for
the relation between the mass of the supermassive black hole in the
center of galaxies and their bulge velocity dispersion (Scannapieco
and Oh 2004, Hopkins 2005, King 2010, Ostriker et al. 2010). They may
also pollute with metals and heat the intergalactic medium, stopping
structure formation.
However, before these outflows can be used as cosmic probes, it is
imperative to understand their nature. Unfortunately, there is
little understanding so far on their origin, geometry and physical
properties. In addition, we still do not know the relation among
the systems with different FWHMs in their absorption lines, the exact relation between the
UV and X-ray absorbers, and any possible connection between the
absorbing and emitting gas in the central regions of AGN. 

It has been suggested that the different UV and X-ray absorption lines
with different FWHMs correspond to a single phenomenon, viewed at
different angles, and
that the same gas may be responsible for absorption and emission
(e.g. Elvis 2000, Krongold et al. 2007, Andrade-Velazquez et
al. 2010). Other ideas suggest that
mini-BALs and BALs may represent the same outflow but at different
evolutionary stages (Hamann \& Sabra 2004). These ideas are not
opposed to each other, as they describe different parts of the
global picture. 

Given that absorption lines provide information on the material
located only in our line of sight to the source, and that the presence
of outflowing material is not always evident in emission lines, one
must rely on a different diagnostic to understand the true nature of these
winds. Line variability has proven to be an effective method to study
AGN winds. However, the variations can be produced by
different  effects, such as changing in the ionization state of the
gas, changes in column density, or changes in covering factor, and
these effects are not always straightforward to differentiate. Line
variations in BALQSOs are generally small, indicating that BAL winds
are long-lived stable flows (Barlow 1994; Lundgren et al 2007; Gibson
et al. 2008). However, in two cases, strong
variations have been found, showing the emergence of BAL (or BAL-like)
outflows. Leighly et al. (2009) reported the appearance of BAL troughs
in the Narrow-Line Seyfert 1 galaxy WVPS 007, an object where a BAL
outflow is not expected given its intrinsic low luminosity. Hamann et
al. (2008) reported the appearance of broad absorption lines in the
spectra of quasar J105400.40+034801.2. The lines, though typical of
BALQSOs, presented a velocity shift of 26,300 km s$^{-1}$, which
resulted in a balnicity index of zero\footnote{The balnicity index
  (which is to be applied to C IV) measures the equivalent width of
  strong absorption features (expressed in km
  s$^{-1}$), but requires that each
  absorption feature contributing to the index spans at least 2000 km
  s$^{-1}$ (to exclude intervening systems) and only includes in the
  calculation absorption regions dipping 10\% or more below the
  normalized continuum. In addition, the first 3000 km
  s$^{-1}$ blueward of the emission peak are excluded to distinguish
  ``associated absorption'' from broad absorption. For further details
  see Appendix A in Weymann et al. (1991).   
}, excluding them from the
standard BAL definition (Weymann et al. 1991).

In this paper we report the emergence of a BAL outflow in the luminous
quasar Ton 34 (PG 1017+280). Ton 34, at $z_q=1.928$, represents an
extreme case of the continuum decline (or UV-break) that takes
place in quasar spectra shortward of $\approx$ 1100 \AA\ (Telfer et
al. 2002, hereafter TZ02), contrasted by a ``normal'' spectrum redward of this
break. If the far UV decline is fitted by a powerlaw (F$_\nu\propto
\nu^{+\alpha}$), the index value in Ton 34 is $\alpha=-5.3$, remarkably
steeper than the `average' spectral energy distribution (SED) derived
by TZ02, which behaves as  $\nu^{-1.76}$ (in fact,
among the 77 far UV indices measured by TZ02, only in 3 objects there
was an ionizing continuum steeper than  $\nu^{-3}$).   
Binette \& Krongold (2008a) suggested that the extreme UV flux might
undergo a recovery shortward of 450 \AA\ (as hinted by IUE data), and
studied the possibility that
the particular shape of the SED could be the result of carbon crystalline dust
absorption (Binette et al. 2005, Haro-Corzo et al. 2007). They found
that moderate columns of dust could indeed reproduce the
UV-break. However, due to the observational limits on atomic gas
absorption, they hypothesized the possibility that the dust was part
of an ionized high-velocity flow in this system. Given the lack of UV
ionizing photons in the SED of Ton 34,
Binette \& Krongold (2008b) further analyzed the emission line
spectrum of this quasar. They find evidence of an unusual strength (relative to
Ly$\alpha$) of low to intermediate excitation lines (such as OII +
OIII $\lambda\lambda 835$, NIII + OIII  $\lambda\lambda 686-703$, and
NIII + NIV $\lambda\lambda 765$), which could not be explained by 
photoionization processes in the emitting gas, but rather by shock
excitation.

\section{Multi-time Spectra of the Quasar TON 34 \label{par:obs}}

The Sloan Digital Sky Survey (SDSS) carried spectroscopic observations
of the
quasar Ton 34 on January 30, 2006. The fully reduced spectrum was retrieved from
the SDSS data release 7 (Abazajian et al. 2009). The spectrum covers
the spectral range
between 3808 and 9215 \AA\ (1335 to 3147 \AA\ in the rest frame of Ton
34), with a resolving power $\sim$2000 (130 km $s^{-1}$). According to
the SDSS data release, an uncertainty of less
than 3\% is expected in the flux calibration. The S/N ratio ranges
from 45 in the blue end to 39 in the red one. A detailed analysis of
the full range spectrum will be presented in a forth coming paper
(Binette et al. in preparation).

To study the spectral variability in Ton 34, we refer to the
published optical spectrum obtained by Sargent et al. (1988). Only for
illustrative purposes, we present here the digitized data presented by
Binette \& Krongold (2008a,b). Briefly, we describe the properties of
the data, as reported by Sargent et al. (1988). The spectrum was taken  at the
Palomar 5.08m Hale Telescope (we focus on the red arm spectrum,
covering the C IV region of interest), between November 19-21, 1981.
The total exposure time was 13,500 s. The resulting spectrum covers
the spectral range between 3600 and 4810 \AA\ (1229 to 1642 \AA\ in
the rest frame of Ton 34) with a resolution $\sim$100 km s$^{-1}$. The
S/N ratio is $\sim$75 in the region between the Ly$\alpha$ and C IV
emission lines. The data was acquired with the sole purpose of
detecting broad and narrow absorption lines, and thus, only a relative
flux calibration was carried out.    

The spectra was corrected for Galactic reddening assuming the Cardelli
et al. (1989) extinction curve corresponding to R$_V$=3.1 and
E$_{B-V}$=0.13. The latter value corresponds to the mean extinction
inferred from the 100$\mu$ maps of Schlegel et al (1998) near Ton
34. We note however, that this correction has no effect on the results
presented here. 
Throughout this paper we assume a cosmology consisting of 
H$_o$=71 km s$^{-1}$ Mpc$^{-1}$, $\Omega_M=0.27$, and
$\Omega_{\lambda}=0.73$.  

\section{The Emergence of a BAL Flow \label{par:analysis}}

\begin{figure}
\plotone{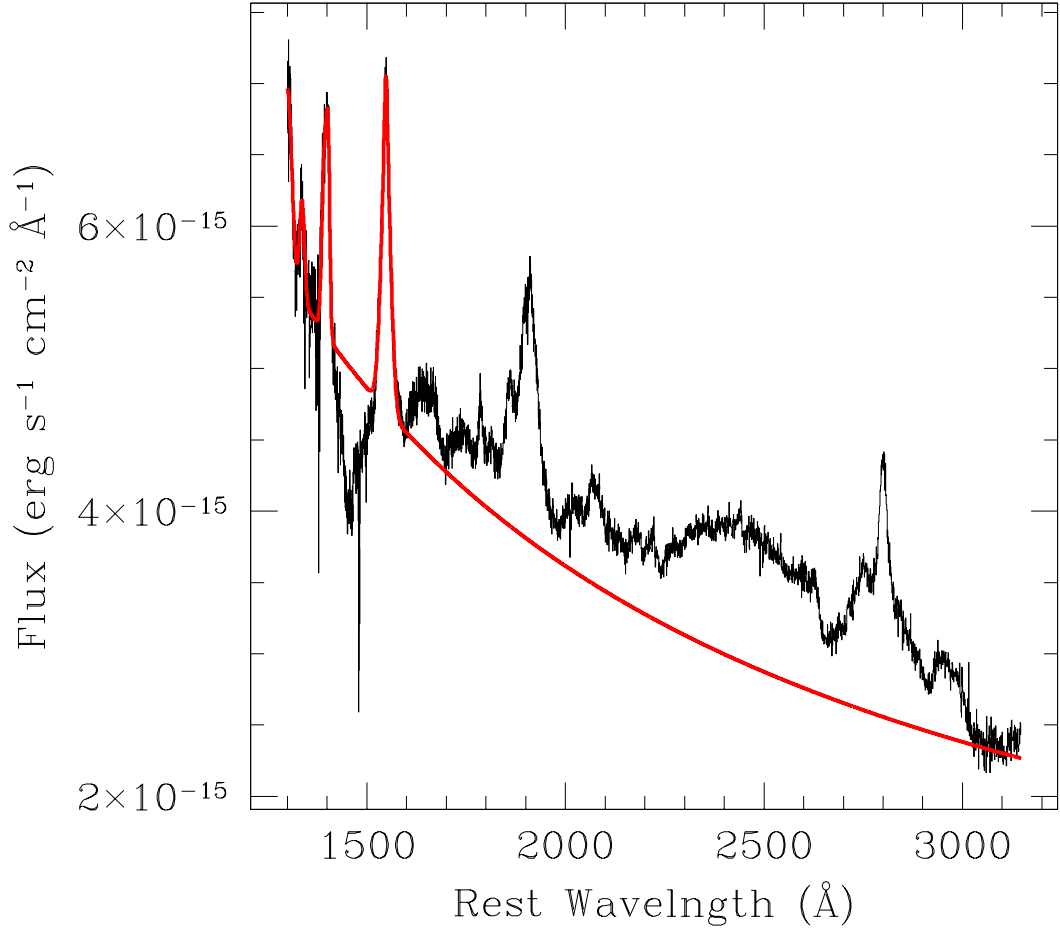} 
\caption[]{Full band rest-frame SDSS spectrum of quasar Ton
  34. 
The red line corresponds to the continuum plus emission lines (in the
blue part of the spectrum) adopted in this paper.   
\label{figure:sloan_fit}}

\end{figure}

Figure \ref{figure:sloan_fit} presents the full band SDSS rest-frame
spectrum of Ton 34. A visual inspection hints at the presence of a
strong broad absorption line blueward of the CIV$\lambda$1549 emission line
region. However, given the numerous emission features present in
the region covered by the spectrum redward of this emission line
(including strong blends of Fe II
transitions), a full band continuum determination is not
straightforward (for a detailed analysis of the full band spectrum see
Binette et al., in preparation). Since our goal is to test the
presence of a CIV BAL trough, we defined the most
conservative continuum level,
as shown in Figure \ref{figure:sloan_fit}. To define this continuum,
we fit a simple powerlaw to the ``line free regions'' in the
full band spectrum, particularly, we forced this simple powerlaw to
fit the regions around 3000-3100 \AA\ (assumed to be a ``good
continuum representation'' given the lack of Fe II emission) and the line   
free regions between 1320 and 1380 \AA\ (so that in the continuum
determination we are assuming no absorption by Si IV). This simple fit
also provides a good continuum description of the two line free regions
close to 1600 \AA\ and 1700 \AA.  
We note that other representations of the continuum (including local
fits to the CIV- SiIV region or broken power laws) are  possible,
though all of them would place the continuum at
higher flux  levels, making the absorption trough stronger. The
emission lines in the blue part of the spectrum  were fit using up
to two different Gaussian components. These lines are
reported in Table \ref{table1}.  

\begin{figure}
\plottwo{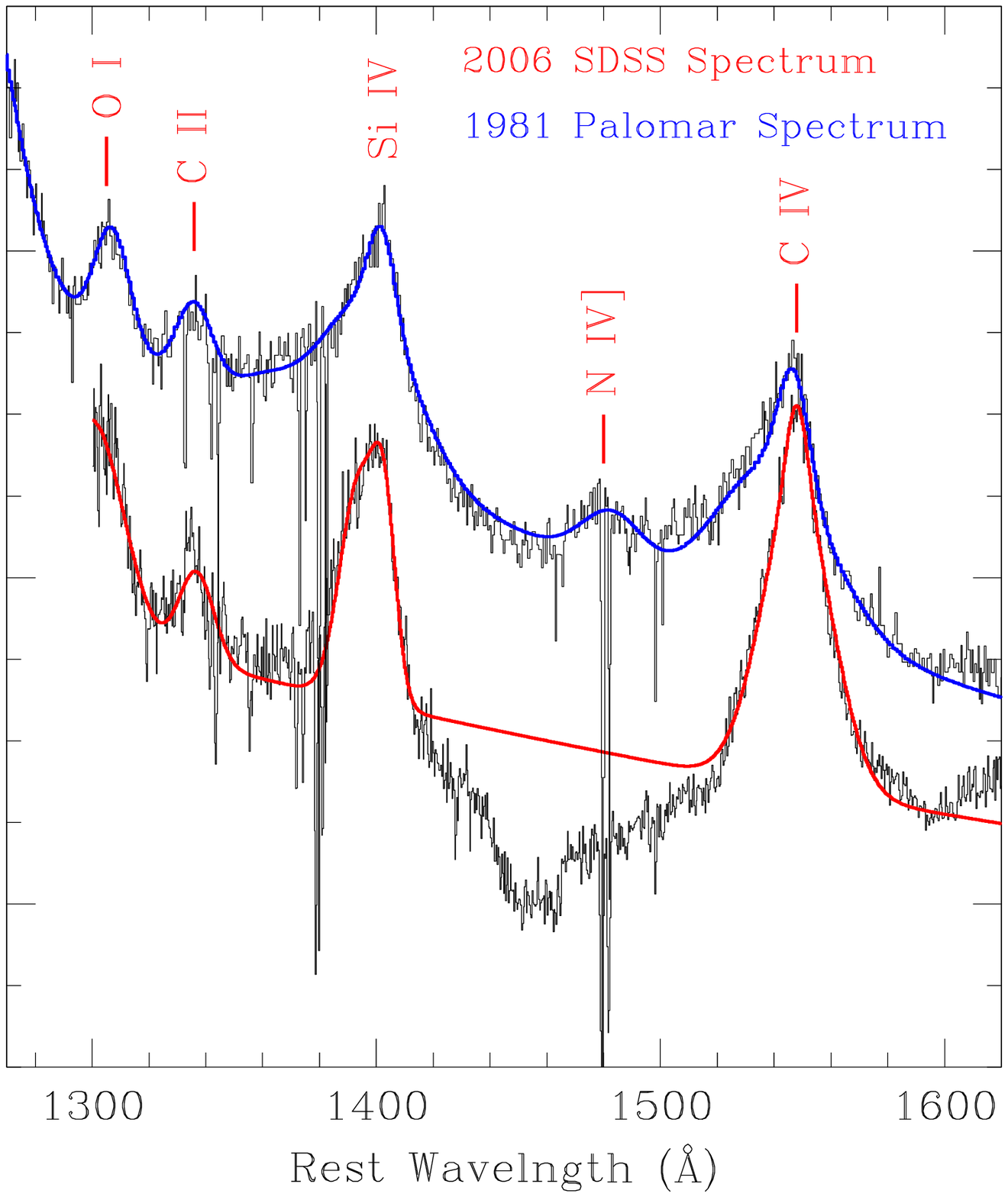}{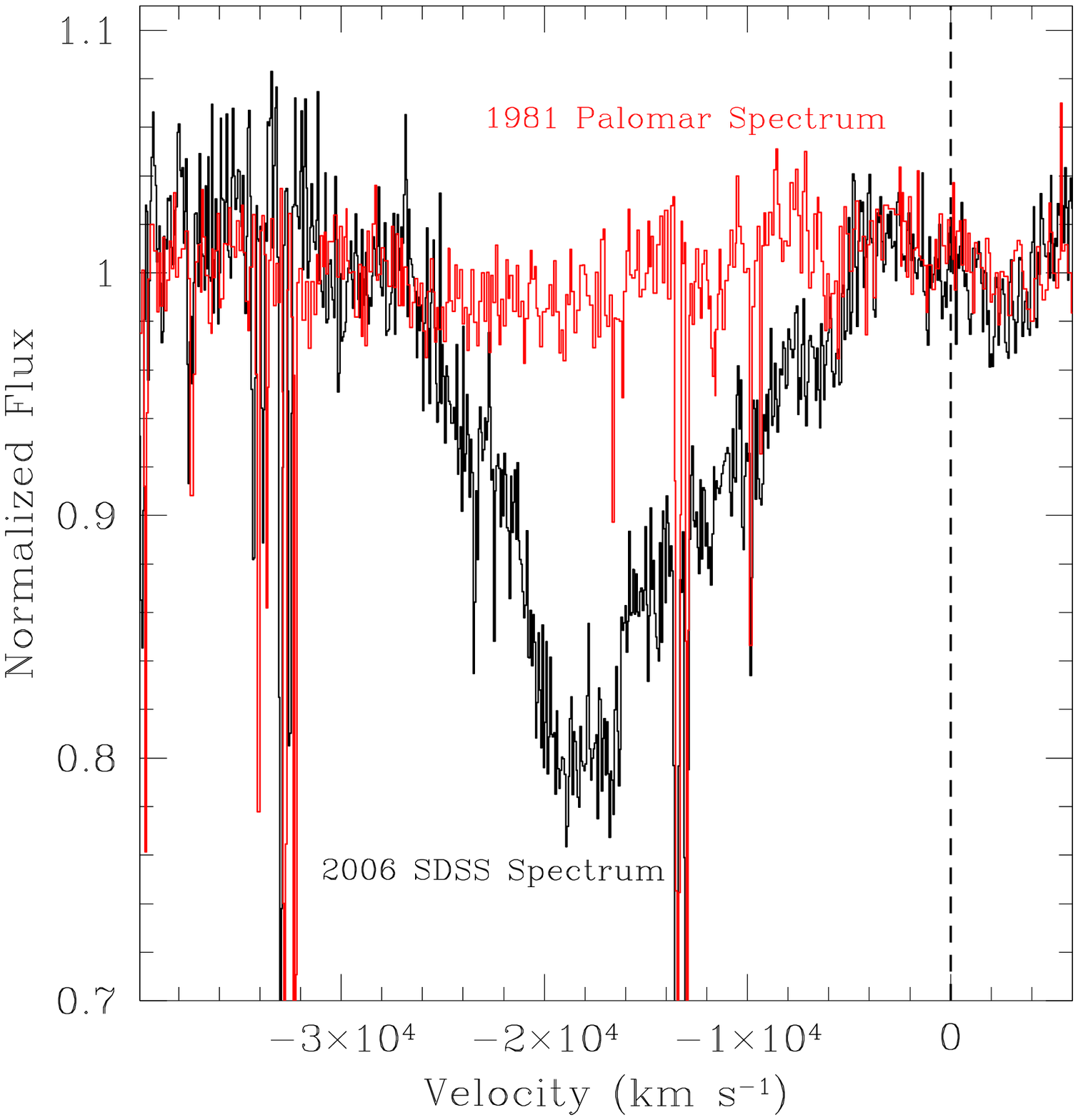} 
\caption[]{Left: 2006 SDSS spectrum of Ton 34 in the 1300-1600 \AA\
  range (rest-frame). The red line represents the (full range)
  continuum plus emission lines fit.  For illustrative purposes, the
  1981 Palomar digitized spectrum (Sargent et al. 1988) is also shown,
  along with the best continuum plus lines fit (blue line). The spectra
  is presented in arbitrary flux units. Right: Normalized SDSS and
  Palomar spectra in velocity space. The emergence of a CIV BAL is evident in the data. 
\label{figure:spectral_fit}}
\end{figure}

Figure \ref{figure:spectral_fit} presents the same fit on the
1300-1600 \AA\ spectral region, along with the normalized spectra (in
velocity space).
The spectral fit to the SDSS spectrum  clearly shows the presence of a
strong broad absorption line consistent with absorption by
CIV$\lambda$1549 with
velocities ranging from $\sim$5,000-25,000 km $s^{-1}$ with respect
to the rest frame of the source. This feature was not present during the
Palomar observation $\approx$24.2 years before ($\approx$8.3 years in
the rest frame). We note, however, that other narrow absorption
features present in the Palomar spectrum (probably intervening systems from the inter-galactic medium) are also present in the SDSS spectrum.
It should be stressed that the Palomar data was obtained
with the sole purpose of discovering absorption lines (both narrow and
broad) in the spectrum
of Ton 34. Yet, no BAL line was reported by Sargent et al
(1988). Additional confirmation of the lack of the BAL comes
from a former spectrum taken at Las Campanas (100 inch telescope) in
March 11, 1981 (Young et
al. 1982), obtained also with the sole purpose of detecting absorption
lines. 

To further illustrate the emergence of the BAL, we present in Figure
\ref{figure:spectral_fit} the earlier Palomar spectrum, digitized. This
plot also shows the Palomar normalized spectrum, after performing a
continuum plus emission line fit, similar to the one applied to the
SDSS data. 
    
\section{Discussion \label{sec:disc}}

The absorption feature present in the SDSS spectrum of Ton 34
resembles typical CIV troughs found in BALQSOs. This feature spans
velocities ranging from $\sim$5,000-25,000 km $s^{-1}$, with a
FWHM$\sim$ 9,000 km s$^{-1}$. Following the standard BAL definition
(Weymann et al., 1991), we calculate a balnicity index $\simgt$600,
but note that this index strongly depends on the continuum
determination, thus, it could easily be much larger. Given our
continuum determination, no absorption by
SiIV was detected, although the possible presence of this line is
strongly dependent on the
actual continuum level. We note however, that SiIV absorption is
not always observed in the spectra of BALQSOs.  

The spectral shape of the absorption feature cannot be adjusted with two Gaussians, shifted by
$\sim 300$ km s$^{-1}$ to account for the CIV$\lambda\lambda$1548,1550
doublet. Rather, a velocity
structure is evident (suggesting a velocity dependent covering
factor), with the absorption becoming more prominent
between 18,000-20,000 km s$^{-1}$. We estimate the column
density directly from the normalized data using equation (9) by
Savage \& Sembach (1991). We integrate the column density over the line profile
assuming the 2:1 ratio in the oscillator strengths of the CIV doublet,
i.e. we assume that the line is optically thin. Doing this we measure
a CIV column density $>2\times10^{15}$ cm$^{-2}$. However, we consider
this a very uncertain
lower limit  given that the lines might be highly saturated and the
covering fraction might be much smaller than one. We have no means to
measure the saturation level of the lines with the current data,
however, saturation and partial covering have been observed in other
UV outflows (e.g. Arav et al. 2001, Hamann \& Sabra 2004, Gabel et
al. 2005). If the line was indeed highly
saturated (optical depth $\tau_o>
3$ in the center of the line), then the CIV column density would be larger
than $10^{16}$ cm$^{-2}$, and from the residual flux in the core of
the line the covering factor would be $\sim$25\%.

Following Hamann et al. (2008) we can compare different
time-scales relevant to the system to the timescale in which the
outflow appeared in the spectrum of Ton 34 ($<$8.3 years, rest
frame of the source), to test different processes responsible for the
emergence of the flow. Assuming that the wind originates in the accretion
disk, then a characteristic time-scale for the flow is simply
t$_{wind}\sim$R$_{wind}$/v$_{wind}$, where R$_{wind}$ is the distance
from the wind to the central source, and v$_{wind}$ the wind velocity.
For accretion-disk winds R$_{wind}$ is expected to lie just beyond the
CIV emission line region (e.g. Everrett 2005, Proga 2007). Using equation (3) from
Kaspi et al. (2007), and the 1350 \AA\ luminosity of Ton 34, we estimate
R$_{CIV}\sim 2\times10^{18}$ cm, which along with
v$_{wind}\sim$19,000 km s$^{-1}$ yields a flow timescale
t$_{wind}\sim$33 yr. This is a factor $\simgt$4 larger than the
elapsed time between the observations, making the hypothesis of 
an arising flow unlikely.

On the other hand, the crossing time of absorbing clouds through the
continuum source is generally smaller than the flow time. A typical crossing
time-scale can be obtained from the ratio between the size of
the emitting region and the transverse velocity of the
clouds, t$_{cross}\sim$R$_{em}$/v$_{tr}$. Assuming v$_{tr}$ is similar
to the rotational speed in the BLR, we can estimate this quantity
from the width of the CIV emission line, thus v$_{tr}\sim$2500 km
s$^{-1}$. Since the size of the emitting region is constrained by the
variability timescale of the source, from variability studies we can
estimate 
R$_{em}$(1500 \AA)$\sim few\times10^{16}$ (e.g. Kaspi et
al. 2007). Thus t$_{cross}\sim$4 yr,
making the possibility that an existing flow got into our line of
sight consistent with the observations.

We further consider the hypothesis that changes in ionization state may
have produced the appearance of the absorption line. We only consider the
possibility of a decrease in ionization as no broad absorption by HI was
observed in the Palomar spectra (Sargent et al. 1988). Thus, if the
wind was already present during the earlier observation, its ionization level
would have to be much larger than during the SDSS one. We do not have accurate
physical properties (i.e. ionization parameter and temperature) of the
absorbing gas to calculate the photoionization equilibrium
time-scale (e.g. Krongold et al. 2005, 2007). However, we can
calculate the recombination time (defined as the inverse of the
recombination coefficient times the electron density of the gas) for
CV, which is
an upper limit to this time-scale (Nicastro et al. 1999). We draw
the recombination coefficient from Shull and Van Steenberg (1982) and
assume a temperature of $few\times10^4$ K. In order for the gas to
recombine and the trough to emerge in the spectrum of Ton 34 in the
8.3 yr interval between the observations, the number density
of the absorber need to
be $>100$ cm$^{-3}$. Thus, the possibility of a decrease in ionization
state is also consistent with the observations. Nevertheless, we
consider this alternate possibility less likely, given that a large change in
the ionizing flux would be required to produce such a large change in
the ionization state of the gas, which is not typically observed in
quasars. On the other hand, the unusual SED of Ton 34 and the lack of
recent extreme UV observations prevents us from ruling out this idea.

Finally, we stress that the extreme far UV break observed in the
spectrum of Ton 34 is difficult to explain in terms of accretion disk
models. This break can be explained (see review by  Binette et al.,
2008c) in terms of atomic absorption by an ionized, nearly relativistic
outflow (Eastman et al. 1983), or in terms of absorption by an
intervening carbon crystalline dust component close to
the quasar. In the case of Ton 34, Binette \& Krongold (2008a) concluded
that the dust component should be part of an ionized medium, and
hypothesized that it could be part of a (moderate velocity) outflow.
If the emission line
region is located closer to the ionization source than this outflow,
this would also bring the required SED by photoionization models to
a much better agreement with the intrinsic SED of the central
source. In the case of three AGN, namely quasar Ton 34, NLSy1
galaxy WVPS 007 (Leighly et al. 2009), and quasar J105400.40+034801.2
(Hamann et al. 2008), the emergence of a BAL outflow has been
reported. For the two quasars the most likely explanation is that the
outflow was already present and simply
intersected our line of sight (also possible for WVPS 007). This may
imply that such outflows are
ubiquitous in quasars, yet only become observable when the absorbing
material accidentally crosses our line vision to the central source.
In the case of Ton 34, a subject that deserves further attention is
the fact that an outflow emerged in a quasar with such an extreme SED.

\acknowledgements 
We thank the anonymous referee for constructive comments that helped to improve
the paper. 
This work was supported by the UNAM
PAPIIT grant IN104009. The results presented in this paper include observations from  the SDSS facilities (obtained from the data archive at the web site
http://www.sdss.org/).


\clearpage
\clearpage
\begin{deluxetable}{lccc}
\tablecolumns{4} \tablewidth{0pc} \tablecaption{Emission Lines in the
  SDSS data  \label{table1}}
\tablehead{ \colhead{Species} & \colhead{$\lambda$ (\AA)} & \colhead{FWHM}
& \colhead{Flux}\\
& & \colhead{(km s$^{-1}$)} &
\colhead{(10$^{-14}$ erg s$^{-1}$ cm$^{-2}$) }} 
\startdata
\cutinhead{Broad Components}
C IV & 1549 & 5600 & 16.0 \\
Si IV & 1402 & 3400 & 7.0 \\
C II & 1335 & 3300 & 2.8 \\
O I & 1302 & 5800 & 6.0 \\
\cutinhead{Narrow Components}
C IV & 1549 & 1700 & 1.4 \\
Si IV & 1402 & 1900 & 2.9 \\
\enddata

\end{deluxetable}

\clearpage

\clearpage


\end{document}